\documentstyle[12pt,epsfig]{article}

\newcommand{\be}{\begin{equation}}
\newcommand{\ee}{\end{equation}}
\newcommand{\bc}{\begin{center}}
\newcommand{\ec}{\end{center}}
\newcommand{\bea}{\begin{eqnarray}}
\newcommand{\eea}{\end{eqnarray}}
\newcommand{\tx}{\textstyle}
\begin{document}

\begin{center}
{\Large \bf  Structure of radiative interferences and g=2 for vector mesons.}\\

\vspace{1.3cm}

{\large  G. Toledo S\'anchez}.\\

{\it  Instituto de F\'{\i}sica, UNAM A.P. 20-364, M\'exico 01000 D.F. M\'exico\\}
{\it Department of Physics, Florida State University, Tallahassee, FL 32306, USA}
\end{center}
\vspace{1.3cm}

\begin{abstract}

The result by Burnett-Kroll (BK) states that for radiative decays the  interference  of ${\cal O}(\omega^{-1})$ in the photon energy $\omega$, vanishes after sum over polarizations of the involved particles. Using radiative decays of vector mesons we show that if the vector meson is polarized the ${\cal O}(\omega^{-1})$ terms are null only for the canonical value of the magnetic dipole moment of the vector meson, namely ${\bf g}=2$ in Bohr's magneton units. A subtle cancellation of all  ${\cal O}(\omega^{-1})$ terms happens when summing over all polarizations to recover the Burnett-Kroll result. We also show the source of these terms and the corresponding cancellation for the unpolarized case and exhibit a global structure that can make them individually vanish in a particular kinematical region.\\
\end{abstract}


\bc
{\large \bf I. Introduction\\}
\ec
 The early work by Low \cite{low}, that relates the radiative process   with the corresponding non-radiative and the electrostatic properties of the involved particles, provided ground  to  develop Bremsstrahlung studies in a model independent basis. These  processes have been used as a way to obtain information of the electromagnetic structure of the particles \cite{zkp}, and the importance of off-shell effects \cite{welsh}.  Subsequent works \cite{zkp,chew,bk} exploited the Low's result to show explicit properties of the so called Low's amplitude. One of these is the work by T. H. Burnett and Norman M. Kroll (BK) \cite{bk} which stated that the interference of the first and second terms of the amplitude expansion, in powers of the photon energy ($\omega$), after sum over polarizations is null, namely the  ${\cal O}(\omega^{-1})$ in the squared amplitude. This  was simultaneously found by V. I. Zakharov et. al.\cite{zkp}. In general those terms in the amplitude can be identified with the electric charge and magnetic dipole emissions respectively and thus the result can be seen as a resemblance of the classical observation of the non-interference of these multipoles. These and other features are embodied in the so-called soft photon approximation.
The subtle cancellation of the interferences in practice is not traced back and thus many interesting features are shadowed by just checking that the results satisfy the Burnett-Kroll theorem. One of these is the existence of additional structures  that can help us  to obtain more information about the decay and the properties of the involved particles.
One can also wonder if the polarized case satisfy the Burnett-Kroll result  and if so, under which conditions. At the same time, it is interesting to know, in practice, how the sum over polarizations leads to the vanishing of $O(\omega^{-1})$ contributions. 
Given the fact that the electric charge is completely determined by charge conservation the magnetic dipole moment (MDM) value naturally plays a key role in radiative decays. For example the $W$ gauge boson MDM is predicted  by the standard model be ${\bf g}=2$ in Bohr's magneton units (we will refer to the MDM by the giromagnetic ratio ${\bf g}$ ) and is a test of the gauge structure of the theory. Indeed, many authors \cite{jackiw} have shown that this value enjoys many interesting features in the description of  electromagnetic phenomena for particles of half and integer spin. Moreover, in theories where the vector mesons are considered as gauge bosons of a hidden symmetry \cite{bando}, the coupling to an electromagnetic field  have  ${\bf g}=2$ in a similar way to the $W$ gauge boson. Thus this particular value is frequently assumed be the canonical one. In this work we offer an additional feature to favor it by observing the radiative decay interferences for polarized vector mesons, there the  $O(\omega^{-1})$ terms are null only if  ${\bf g}=2$.
 By another hand, in the soft-photon approximation, was observed \cite{zkp,us97,usprd} that for radiative decays involving vector mesons is possible to suppress the electric charge contribution to the photon energy spectrum by an appropriate choice of a kinematical configuration. In a most recent work \cite{usiop} an exploration beyond that approximation showed that  {\it the interference of the electric charge radiation with any gauge invariant term} of the transition amplitude exhibit a typical structure that in particular is null by the same kinematical configuration, namely that where the photon is collinearly emitted off the charged particle of the final state when in the initial particle restframe. Although not explicitly mentioned this is also true for each of the $O(\omega^{-1})$ terms whose total sum is null as stated by Burnett and Kroll. Here we apply that result to unpolarized vector mesons radiative decays to explicitly exhibit this behavior and how the BK result is gotten, drawing an insight into the destructive interferences. All over the interference contributions the MDM value is involved and therefore its role can be inferred. These are the  questions that we plan to address along the paper.\\

In the present work we explore all this for radiative decays of vector mesons  typically of the form $V^+ \rightarrow P^+ P'^0 \gamma$, where $V$ ($P,P'$) is a vector (pseudoscalar) meson.
 We start by  considering the vector meson polarization to study the effect of the MDM in the  radiation probability structure, then we state the features of the interferences in the unpolarized case and there we exhibit how the BK result is obtained. At the end we discuss  our results and their implications.\\

\bc
{\large \bf II. Polarized radiation distribution\\}
\ec

To make clear the structure of the Burnet-Kroll terms in the following we  write the explicit gauge invariant Low's amplitude for the decay of a vector meson that we choose be $\rho^+ \rightarrow
\pi^+  \pi^0  \gamma$, although the results are not restricted to this particular case. We will use the 4-momentum  notation $q$ $\rightarrow $ $p$ $p'$ $k$ in the respective order and $\epsilon$ ($\eta$) correspond to the polarization 4-vector of the photon (vector meson). Thus, the Low's amplitude can be written as \cite{us97}:

\begin{eqnarray}
{\cal M}_L&=&ieg_{\rho \pi\pi} \left[
 (p-p')\cdot\eta L\cdot \epsilon^*+ L\cdot \epsilon^* k\cdot \eta
-\frac{{\bf g} p\cdot k}{q\cdot k}\left(\frac{p\cdot\epsilon^*}{p\cdot k}k\cdot\eta-\epsilon^*\cdot\eta \right) \right. \nonumber\\
&&\left. +\left(2-\left(1-\frac{{\bf g}}{2}\right)\left(1+\frac{\Delta^2}{m_\rho^2}\right)\right)\left( \frac{q\cdot\epsilon^*}{q\cdot
k}k\cdot\eta-\epsilon^*\cdot\eta \right)
 \right]\label{lowamplitude}
, 
\end{eqnarray}

\noindent where  $g_{\rho \pi\pi}$ denotes the $ \rho \pi\pi$ coupling, $e$ is the electric charge of the positron, $m_\rho$ is the $\rho$ meson mass and $m_\pi$ ($m_{\pi^0}$) is the mass of the charged (neutral) pion. The magnetic dipole moment is given by ${\bf g}$ in Bohr's magneton units, $\Delta^2 \equiv m^2_\pi -m^2_{\pi^0}$ and $L^\mu
 \equiv \left( \frac{p^\mu}{p\cdot k} - \frac{q^\mu}{q\cdot k}\right)
$.\\

Let us now study the interferences behavior by considering the polarization of the vector meson. The magnetic dipole moment ${\bf g}$ is not restricted to a particular value and the masses of the charged ($m_\pi$) and neutral ($m_{\pi^0}$) pions are different. We consider, for simplicity, the restframe of the decaying particle and the radiation gauge ($\epsilon_0=0$). The first condition imply that the vector meson polarization tensor $\eta$ has the form $ \eta^{(j)}_\mu=(0,\vec{\eta}^{(j)}) $with $j$ = 1, 2, 3. For definiteness we choose the following base:

\[
\vec\eta^{(1)}=\frac{1}{\sqrt{2}}(1,i,0)
\]

\be
\vec\eta^{(2)}=\frac{1}{\sqrt{2}}(1,-i,0) \label{pola}
\ee

\[
\vec\eta^{(3)}=(0,0,1)
\]

The coordinate system is taken in such a way that the photon vector momentum is along the $\hat z$ direction, this means  $k=(\omega,0,0,\omega)$ which, in the gauge radiation ,implies that $\vec k\cdot\vec\epsilon = 
\omega \epsilon_3 = 0$ and therefore the photon polarization tensor $\epsilon_\mu$ can be written as:

\[
\epsilon_\mu = (0,\epsilon_1 ,\epsilon_2 ,0).
\]

Once we have established our conventions we proceed to compute the polarized amplitudes from equation (\ref{lowamplitude}). The explicit expressions after simplifications, for each of the vector meson polarizations  (\ref{pola}) are the following, respectively:

\be
{\cal M}^{(1)}=ie\frac{g_{\rho\pi\pi}}{\sqrt{2}}\left[ \frac{{\tx {\bf g}}p\cdot k}{{\tx q\cdot k}}-2+
\left(1+\frac{\Delta^2}{m^2_\rho}\right)
\left(1-\frac{{\tx {\bf g}}}{2}\right)\right] (\epsilon^*_1 +i\epsilon^*_2) 
+\sqrt{2}ieg_{\rho\pi\pi}\frac{{\tx p\cdot\epsilon^*}}{{\tx p\cdot 
k}}(p_1+ip_2)
\ee

\be
{\cal M}^{(2)}=ie\frac{g_{\rho\pi\pi}}{\sqrt{2}}\left[ \frac{{\tx {\bf g}}p\cdot k}{{\tx q\cdot k}}-2+
\left(1+\frac{\Delta^2}{m^2_\rho}\right)
\left(1-\frac{{\tx {\bf g}}}{2}\right)\right] (\epsilon^*_1 -i\epsilon^*_2) 
+\sqrt{2}ieg_{\rho\pi\pi}\frac{{\tx p\cdot\epsilon^*}}{{\tx p\cdot 
k}}(p_1-ip_2)
\ee

\be
{\cal M}^{(3)}= 2ieg_{\rho\pi\pi}\frac{{\tx p\cdot\epsilon^*}}{{\tx p\cdot k}}
\left[p_3+\omega\left(1-\frac{{\bf g}p\cdot k}{2 q\cdot k}\right)
\right]
\ee

\vskip0.5cm
In a 3 body decay, besides the masses, only 2 Lorentz  invariants are independent we choose them be $p\cdot k$ and $q\cdot k$ to exhibit the dependence on the photon energy. The total probability transition for each of the 3 directions of the polarization can then be computed and expressed in terms of those kinematical variables as follows:

{\samepage

\bea
\mid {\cal M}^{(1)}\mid^2 
=&\left(\tx eg_{\rho\pi\pi}\frac{\tx p\cdot\epsilon^*}{\tx p\cdot k}\right)^2 &
\left[ 2m^2_{\rho}\frac{p\cdot k}{q\cdot k}
\left(
1+\frac{\Delta^2}{m^2_{\rho}}-\frac{p\cdot k}{q\cdot k}\right) -2m^2_\pi \right.\nonumber\\
&&\left. - 2p\cdot k \left(1-\frac{{\bf g}}{2}\right)
\left( 1+\frac{\Delta^2}{m^2_{\rho}}-2\frac{p\cdot k}{q\cdot k}\right) 
\right]\nonumber\\ 
&-\frac{\tx (eg_{\rho\pi\pi})^2}{\tx 2}{\tx \epsilon^*\cdot\epsilon} &
\left[{\bf g}\frac{p\cdot k}{q\cdot k}-2+\left(1+\frac{\Delta^2}{m^2_{\rho}}\right)\left(1-\frac{{\bf g}}{2}\right)
\right]^2,\nonumber\label{ap1}
\eea
}

\be
\mid {\cal M}^{(2)}\mid^2 = \mid {\cal M}^{(1)}\mid^2 \label{ap3}
\ee

\bea
\mid {\cal M}^{(3)}\mid^2 = & \left(\tx eg_{\rho\pi\pi} \frac{\tx p\cdot\epsilon^*}{\tx p\cdot k} \right)^2 &\left[
m^2_{\rho}\left( 1+\frac{\Delta^2}{m^2_{\rho}}-2\frac{p\cdot k}{q\cdot k}\right)^2
+ 4\left(\frac{p\cdot k}{m_{\rho}}\right)^2 \left(1-\frac{{\bf g}}{2}\right)^2 \right.\nonumber\\
&&\left. +4 p\cdot k \left(1-\frac{{\bf g}}{2}\right)
 \left(1+\frac{\Delta^2}{m^2_{\rho}}-2\frac{p\cdot k}{q\cdot 
k}\right)
\right].\nonumber
\eea
\\

These equations exhibit clearly the structure $\mid {\cal M}^{(i)} \mid^2 = \frac{A^{(i)}}{\omega^2}+\frac{B^{(i)}}{\omega}+C^{(i)}\omega^0$, and there can be observed that the  ${\cal O}(\omega^{-1})$ terms are no null unless the magnetic dipole moment  takes the canonical value ${\bf g}=2$. This result is very interesting because this simplifying-feature was also found in reference \cite{BMT} when describing the equation of motion  of the polarization tensor for a particle in a homogeneous external electromagnetic field and reinforces the observation by many authors \cite{jackiw} that the choice ${\bf g}=2$ for charged vector mesons leads to richer properties of the radiative process description.
We can also notice that the  ${\cal O}(\omega^{-1})$ terms are proportional to the kinematical factor $(\frac{p\cdot\epsilon^*}{p\cdot k})^2$ whose importance will be clarified in the next section.\\

\noindent Adding the  three  polarized equations (\ref{ap3}) we render to the unpolarized case

{\samepage
\begin{eqnarray}
 \sum_\eta \mid {\cal M}\mid^2& = & (eg_{\rho\pi\pi})^2\left( 
\frac{p\cdot\epsilon^*}{p\cdot k}\right)^2 \left[
\left(1+\frac{\Delta^2}{m^2_{\rho}}\right)^2 m^2_{\rho}-4 m^2_\pi
+4\left(\frac{p\cdot k}{m_{\rho}}\right)^2
\left(1-\frac{{\bf g}}{2}\right)^2
\right]\nonumber\\
&&-(eg_{\rho\pi\pi})^2\epsilon^*\cdot\epsilon
\left[{\bf g}\frac{p\cdot k}{q\cdot k}-2+\left(1+\frac{\Delta^2}{m^2_{\rho}}\right)\left(1-\frac{{\bf g}}{2}\right)
\right]^2
,\label{apf}
\end{eqnarray}
}

\noindent which is free of ${\cal O}(\omega^{-1})$ terms, independently of the MDM value, in accordance with the BK theorem.\\

\bc
{\large \bf III. Unpolarized radiation distribution\\}
\ec

In a recent work \cite{usiop} we  showed that in a radiative 3 body decay involving vector mesons the interference between  the electric charge and gauge invariant terms have a typical structure regardless the order in the photon energy. This is a result of two features of the total  amplitude, which can be written as follows:
\be
{\cal M}\propto \epsilon^* \cdot L + M\cdot \epsilon^* 
\ee
i).- {\it The radiation from electric charges is gauge invariant through the tensor $L^\mu
 \equiv \left( \frac{p^\mu}{p\cdot k} - \frac{q^\mu}{q\cdot k}\right)
$.} with $q$ and $p$  the four-momenta of the initial and final charged particles, respectively, and $k\ (\epsilon)$ the four-momentum (polarization)  of the photon.\\
ii).-{\it The electromagnetic gauge invariance  of the other terms} summarized in $M$, which can be  of any order in the photon energy.\\

\noindent After sum over the vector meson polarizations, the interference between these two terms can be written as \cite{usiop}
\be
\sum_{\eta}  (L \cdot \epsilon^* ) (\epsilon \cdot M) = a_2 (L\cdot \epsilon^*)^2 .\label{uinterference}
\ee

\noindent with $a_2$ a Lorentz invariant function. The corresponding limit for the soft photon approximation is the Burnett-Kroll term, which must be null after sum of {\it all}  ${\cal O} (\omega^{-1})$ contributions i.e. because $(L\cdot \epsilon^*)^2$ is of ${\cal O} (\omega^{-2})$ then the ${\cal O}(\omega)$ term from $a_2$ does not contribute to the squared amplitude. Eqn. (\ref{uinterference}) suggests that kinematical advantages can be gotten from considering the restframe of the decaying particle, there, after sum over the photon polarizations, we have that
\be 
L^2=L \cdot L =-\frac{|\vec{p}|^2}{(p\cdot
k)^2} \sin^2\theta
\ee
\noindent with $\theta$ the angle between the photon three-momentum and 
$\vec{p}$. Thus the interference shows an important angular dependence,  namely the cancellations are less crucial as we approach to the collinear case up to  become null and conversely.
The  ${\cal O}(\omega)$ part of the $a_2$ function should vanishes as prescribed by BK and therefore a game between the two conditions can be played.\\

In the following we show this structure for the decay under consideration ($\rho \rightarrow \pi\pi\gamma$) and how  the BK cancellation happens. For the sake of clarity in this section we assume isospin symmetry $(m_\pi=m_{\pi^0})$ and the magnetic dipole moment be ${\bf g}= 2$, in Bohr's magneton units. Upon this conditions in the Low's amplitude (\ref{lowamplitude}), we identify the contributions.
The first term of the amplitude is of ${\cal O}(\omega^{-1})$ (here after ${\cal M}_e$) and can be identified with the electric charge radiation. The remaining terms which are
of order $\omega^0$  are explicitly gauge invariant
. We will refer to them as  ${\cal M}_0$.
The interference of these gauge invariant amplitudes are summarized in Table 1. We have included the  result of the  electric charge interference with itself\footnote{ Examples of higher order interferences can be found at \cite{usiop}.},and the corresponding for  ${\cal M}_0$  because the cancellation of the Burnett-Kroll terms are linked  as we show below.\\
\begin{center}
%{\large
\begin{tabular}{|c|ll|}
\hline
\hline
&&\\
 $\sum_{\eta}{\cal M}_e{\cal M^*}_e$ &$ (eg_{\rho\pi\pi} L\cdot \epsilon)^2 [m_\rho^2-4m_\pi^2 $&$ +\frac{1}{m_\rho^2}(2p\cdot k- q\cdot k)^2-2q\cdot k]$ \\
&&\\
\hline
&&\\
 $2Re\sum_{\eta}{\cal M}_e {\cal M}_0^* $&$  (eg_{\rho\pi\pi} L\cdot \epsilon)^2 [$&$ -\frac{2}{m_\rho^2}(2p\cdot k-q\cdot k)^2+2q\cdot k]$ \\
&&\\
\hline
&&\\
 $\sum_{\eta}{\cal M}_0{\cal M^*}_0$&$- 4(eg_{\rho\pi\pi})^2\epsilon\cdot \epsilon^*(1-\frac{p\cdot k}{q\cdot k})^2$ &$+ (eg_{\rho\pi\pi}L\cdot \epsilon)^2 \frac{1}{m_\rho^2}(2p\cdot k-q\cdot k)^2 $ \\
&&\\
\hline
\hline
&&\\
 $\sum_{\eta}{\cal M}_L{\cal M^*}_L$& $ (eg_{\rho\pi\pi} L\cdot \epsilon)^2(m_\rho^2-4m_\pi^2)$&$- 4(eg_{\rho\pi\pi})^2\epsilon\cdot \epsilon^*(1-\frac{p\cdot k}{q\cdot k})^2 $ \\
&&\\
\hline
\hline
\end{tabular}
%}

Table 1. Interferences  for the decay $ \rho^+  \rightarrow
\pi^+  \pi^0  \gamma$ \\
\end{center}

It is interesting to note that the cancellation of  ${\cal O}(\omega^{-1})$ terms is a result of a combination between the electric charge radiation itself, the interference term and the non-electric one.
 After sum over the photon polarization,  owing to $L^2$, their proportionality to $sin^2(\theta)$ ( $\theta$ the angle between the photon and  charged pion 3-momenta) is a global factor and although the  ${\cal O}(\omega^{-1})$  terms vanish after summation, they individually do not do so and therefore can be used to understand how important these cancellations are, in particular can be seen that they are suppressed until vanish as we approach to the collinear case and conversely.\\
 It is straightforward to show that the result in the last line of the table coincides with the unpolarized case (\ref{apf}) obtained previously, with ${\bf g}=2$ and $\Delta^2=0$.\\

Let us now show how  their relative magnitudes are in the photon energy angular distribution. Figure 1 shows the photon energy spectrum of each term for an angle $\theta=10^0$. The long-dashed, short-dashed and light solid lines are $\sum_{\eta}{\cal M}_e{\cal M^*}_e$, $2Re\sum_{\eta}{\cal M}_e {\cal M}_0^* $ and  $\sum_{\eta}{\cal M}_0{\cal M^*}_0$   respectively. The bold solid line correspond to the total spectrum. We can observe that for a long part  of the region the interference contribution surpass  the total spectrum, signing so its relevance. This being crucial in the final region. As expected, the electric radiation is the leading one at low photon energies.\\

\bc
{\large \bf IV. Discussion\\}
\ec

We  have computed the transition probability for the process $\rho^+ \rightarrow \pi^+\pi^0 \gamma$, as a typical radiative vector meson decay, by considering each of the 3 degrees of polarization of the vector meson. We found that at difference of BK result, the ${\cal O}(\omega^{-1})$ contributions, which partially come from the interference between the electric charge and the magnetic dipole moment radiation, are not null for each polarization state, unless the MDM takes the canonical value ${\bf g}=2$. This result is important if we consider that no further restriction on its value is required elsewhere and that this one is the same that has been found for other authors leads to particular properties of the radiative processes. The presence of   ${\cal O}(\omega^{-1})$ terms is due to the effect of the vector meson spin and disappear after sum over polarizations, satisfying the BK theorem. Our results are important in the sense that qualitatively different behavior arises for the radiation as a consequence of the electromagnetic properties of the vector mesons.\\
In addition, we  studied the  interferences  of the unpolarized radiative decay amplitude of vector mesons, using the fact that the gauge invariance requirements for the amplitude  yields a typical structure to the interferences between the charge and gauge invariant terms. 
In particular we showed this for the  interference with the magnetic contribution, which as stated by Burnett-Kroll vanishes after summation of all  ${\cal O}(\omega^{-1})$ terms of the squared  amplitude. There we  exhibit the subtle cancellation  and exploited the  kinematical facility coming from the global factor $L^2$ to explore the importance of the interferences by looking their relative magnitudes.  In our analysis we have only exploited the gauge invariance  properties of the contributions and the known form of the radiation off electric charges.

Finally we want to point out that the commonly shadowed radiation interferences can offer further interesting information about the particle properties.\\

{\bf {\large  Acknowledgments}}. The author is grateful to G. L\'opez Castro for useful conversations and acknowledge financial support from Conacyt.

\newpage
\begin{figure}
\label{BK10}
\centerline{\epsfig{file=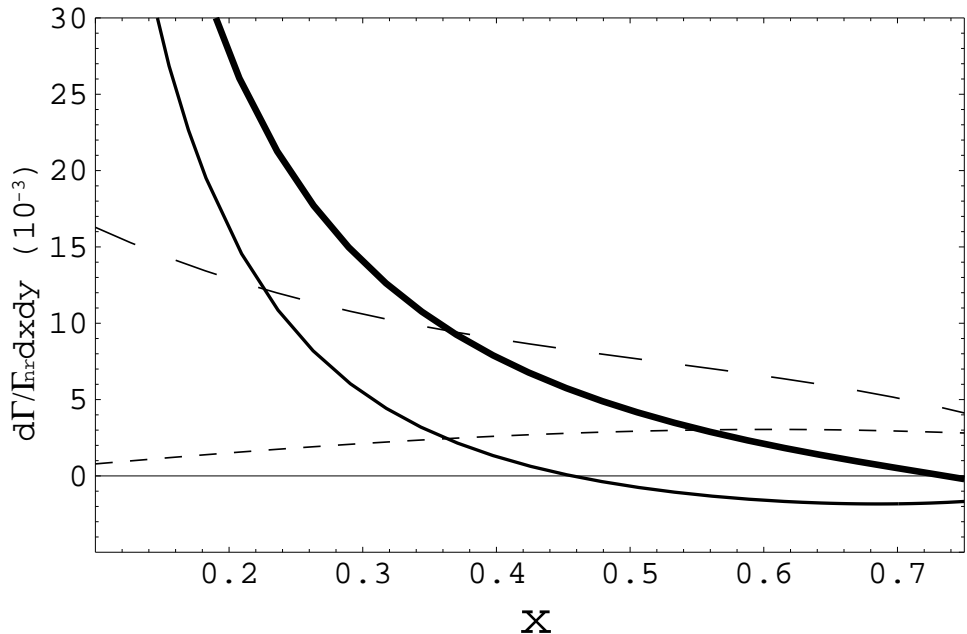,angle=0,width=5in}}
\vspace{-0.1in}
\caption{ Angular and energy distributions of photons in the
process $\rho^+ \rightarrow \pi^+\pi^0 \gamma$, normalized to the
non-radiative rate, as a function of the photon
energy ($X \equiv 2\omega/m_\rho$) for an angle $\theta=10^0$ ($y=cos\theta$) between the photon and the charged pion 3-momenta. The long-dashed, short-dashed and light solid line are  $\sum_{\eta}{\cal M}_e{\cal M^*}_e$, $2Re\sum_{\eta}{\cal M}_e {\cal M}_0^* $ and  $\sum_{\eta}{\cal M}_0{\cal M^*}_0$   respectively. The bold solid line
correspond to the total spectrum.}

\end{figure}

\end{document}